\documentclass[rnote]{aa} 

\usepackage{graphicx}
\usepackage{caption}
\usepackage{subcaption}
\usepackage{txfonts}
\usepackage{natbib}
\usepackage{url}
\usepackage{amsmath}
\begin{document}

   \title{Homogeneous spectroscopic parameters for bright planet host stars from the northern 
   hemisphere\thanks{Based on observations obtained at the Telescope Bernard Lyot (USR5026) operated by the 
   Observatoire Midi-Pyr\'en\'ees and the Institut National des Science de l'Univers of the Centre National 
   de la Recherche Scientifique of France (Run ID L131N11 - OPTICON\_2013A\_027).}
   }
   \subtitle{The impact on  stellar and planetary mass}
   \author{S. G. Sousa\inst{1,}\inst{2}
          \and N. C. Santos\inst{1,}\inst{2}
          \and A. Mortier\inst{1,}\inst{3}
          \and M. Tsantaki\inst{1,}\inst{2}
          \and V. Adibekyan\inst{1}
          \and E. Delgado Mena\inst{1}
          \and G. Israelian\inst{4,}\inst{5}
          \and B. Rojas-Ayala\inst{1}
          \and V. Neves\inst{6}
          }

          \institute{
          Instituto de Astrofísica e Ciências do Espaço, Universidade do Porto, CAUP, Rua das Estrelas, PT4150-762 Porto, Portugal
          \and Departamento de F\'isica e Astronomia, Faculdade de Ci\^encias, Universidade do Porto, Rua do Campo Alegre, 4169-007 Porto, Portugal
          \and SUPA, School of Physics and Astronomy, University of St Andrews, St Andrews KY16 9SS, UK
          \and Instituto de Astrof\'isica de Canarias, 38200 La Laguna, Tenerife, Spain
          \and Departamento de Astrof\'isica, Universidade de La Laguna, E-38205 La Laguna, Tenerife, Spain
          \and Departamento de Física, Universidade Federal do Rio Grande do Norte, Brazil
             }

   \date{Received September 15, 1996; accepted March 16, 1997}

 
  \abstract
  {}
   {In this work we derive new precise and homogeneous parameters for 37 stars with planets. For this purpose, we analyze high 
   resolution spectra obtained by the NARVAL spectrograph for a sample composed of bright planet host stars in the northern hemisphere. The new parameters 
   are included in the SWEET-Cat online catalogue.}
   {To ensure that the catalogue is homogeneous, we use our standard spectroscopic analysis procedure, ARES+MOOG, to derive effective temperatures, 
    surface gravities, and metallicities. These spectroscopic stellar parameters are then used as input to compute the stellar mass and radius, which are 
    fundamental for the derivation of the planetary mass and radius.}
   {We show that the spectroscopic parameters, masses, and radii are generally in good agreement with the values available in online databases of 
   exoplanets. There are some exceptions, especially for the evolved stars. These are analyzed in detail focusing on the effect of the stellar 
   mass on the derived planetary mass.
   }
   {We conclude that the stellar mass estimations for giant stars should be managed with extreme caution when using them to compute the planetary masses. 
   We report examples within this sample where the differences in planetary mass can be as high as 100\% in the most extreme cases.}

   \keywords{planetary systems -- stars: solar-type -- stars: abundances -- catalogs}

   \maketitle
%

\section{Introduction}

The stellar characterization of planet hosts is fundamental and has a direct impact on the derivation of the bulk properties of exoplanets. Moreover, it  provides unique 
evidence for the understanding of planet formation and evolution. One early indication was  that the giant exoplanets were preferentially 
discovered orbiting metal-rich stars \citep[][]{Gonzalez-1997, Santos-2000a, Santos-2004b, Fischer_Valenti-2005, Sousa-2008}. Later, it was also
observed that this metallicity correlation is not the same for less massive planets \citep[][]{Sousa-2011b, Mayor-2011, Buchhave-2012}. This evidence 
give strength to the theory of core-accretion for the planet formation \citep[e.g.,][]{Mordasini-2009,Mordasini-2012}. We note, however, that
although these correlations seem to be clear for dwarf stars, the same scenario is not so clear for evolved giant stars \citep[e.g.,][]{Maldonado-2013, Mortier-2013}. 
Interestingly, the stellar metallicity  plays a crucial role not only on planet formation, but also on the evolution and architecture of the planetary systems. For 
instance, \citet[][]{Beauge-2013} and \citet[][]{Adibekyan-2013} found that planets around metal-poor stars show longer periods, and probably migrate less.

These are just a few of the many examples revealing the importance of  stellar characterization for the study of planetary systems. One fundamental 
aspect common
to many of these studies is the homogeneity of methods used to characterize the stars. With this in mind, the SWEET-Cat (Santos et al. 2013) catalogue 
is available to the community with a very ambitious objective: to provide homogeneous spectroscopic parameters for all planet hosts detected with 
radial velocity, astrometry, and transit techniques. In this work we focus our attention on the bright northern hemisphere targets that were significantly 
lacking in this catalogue, and therefore we increase the number of these planet hosts with homogeneous spectroscopic 
parameters (84\% to 93\% for all known RV planet host stars known at the present time).


\section{Data}

The stars analyzed in this work were selected in order to extend the SWEET-CAT catalogue with missing homogeneous parameters. In order to save 
telescope time, we  only focused on northern brighter planet host stars (V < 9 and $\delta > +30º$) for 
which there were not any suitable spectra available in any high resolution spectrograph archive.

The spectroscopic data were collected between  16 April 2013 and  20 August 2013 with the NARVAL spectrograph located at 
the 2-meter Bernard Lyot Telescope (@ Pic du Midi). The data was obtained through the Opticon proposal (OPTICON\_2013A\_027). The spectra were 
collected in ``spectroscopic/object only'' mode which allows us to have a high resolution of R $\sim$ 80\ 000. The exposure 
times for the stars were chosen in order to reach high signal-to-noise ratio for all the targets ($\geq$ 200). The spectra were reduced 
using the data reduction software Libre-ESpRIT \citep[][]{Donati-1997} from IRAP (Observatoire Midi-Pyrénées).

The star \object{HD 132563B} was observed at two epochs with low signal-to-noise. These two spectra were combined in order to increase the 
signal-to-noise for this star. This was achieved using the \textit{scombine} routine which is part of IRAF \footnote{IRAF is distributed 
by National Optical Astronomy Observatories, operated by the Association of Universities for Research 
in Astronomy, Inc., under contract with the National Science Foundation, U.S.A.}.

\section{Spectroscopic parameters}

\subsection{Spectroscopic analysis}

To derive homogeneous spectroscopic parameters for SWEET-Cat, we made use of our standard spectroscopic analysis 
\citep[ARES+MOOG; see][]{Sousa-2008, Molenda-2013, Sousa-2014}. In summary, the measurement of the equivalent widths (EWs) was done automatically and 
systematically using the ARES code \citep[][]{Sousa-2007}\footnote{\url{http://www.astro.up.pt/~sousasag/ares}}. The EWs were then used together with a grid of 
Kurucz Atlas 9 plane-parallel model atmospheres \citep[][]{Kurucz-1993}. The abundances were computed using MOOG\footnote{The 
source code of MOOG can be downloaded at \url{http://verdi.as.utexas.edu/moog.html}}. For the analysis, we first used the linelist 
from \citet[][]{Sousa-2008}. However, since a large number of the planet hosts turned out to be cool K stars, 
we reanalyzed the stars with temperatures lower than 5200 K with a recent linelist from \citet[][]{Tsantaki-2013}.

During the analysis we identified three planet hosts for which we were not able to derive reliable parameters with ARES+MOOG. Two of them (\object{HD 8673} 
and \object{XO-3}) revealed a significant high rotational velocity, making the measurement of the individual equivalent widths difficult. For these cases the 
use of a new analysis based on a synthesis method is presented in \citet[][]{Tsantaki-2014}.\footnote{The 
parameters presented in \citet[][]{Tsantaki-2014} for \object{XO-3} are Teff = 6781 $\pm$ 44 K, log g = 4.23 $\pm$ 0.15 dex, and [Fe/H] = -0.08 $\pm$ 0.04 dex. These 
were actually derived for spectra with higher signal-to-noise observed with the SOPHIE spectrograph. Using the NARVAL spectra we also obtain consistent 
parameters: Teff = 6730 $\pm$ 44 K, log g = 4.33 $\pm$ 0.15 dex, and [Fe/H] = -0.04 $\pm$ 0.08.}

The third planet host for which we were not able to have reliable results with ARES+MOOG is \object{HD 208527}, marked as a K5V star in SIMBAD, but which is actually 
an M giant star \citep[][]{Lee-2013}. The problem is also connected with the difficulty of measuring equivalent widths. Here we also have strongly blended lines 
due to the presence of many more spectral lines, including those from molecular species. 

Table \ref{tab2} presents the spectroscopic parameters derived with ARES+MOOG. The errors were determined following the 
same procedure as in previous works \citep[][]{Santos-2004b, Sousa-2008, Sousa-2011b}.

We would also like to note that three stars in our sample were already analyzed in the SWEET-Cat
catalogue with the same procedure but using other spectroscopic data, allowing us to check the consistency of our results. 
The stars \object{HD 118203} and \object{HD 16175} 
were analyzed in \citet[][]{Santos-2013} using spectra from SARG and FIES, respectively, while \object{HD 222155} was analyzed in the 
work of \citet[][]{Boisse-2012} using spectra from the SOPHIE spectrograph. Our results are consistent with those previously obtained with other spectrographs.

\begin{figure*}[ht]
  \centering
  \includegraphics[width=16cm]{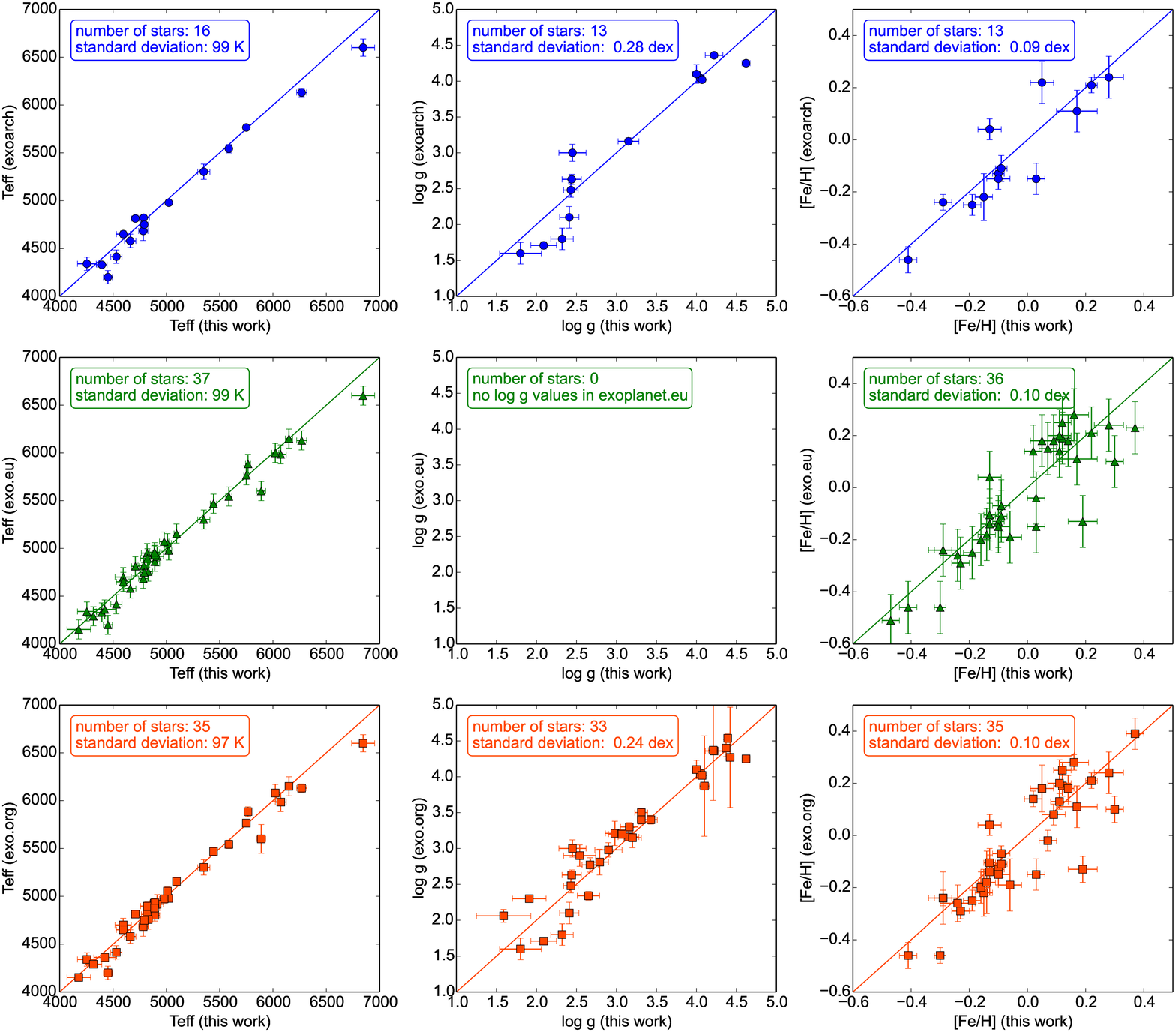}
  \caption{Comparison of the spectroscopic stellar parameters (Teff - left  panels; log g - middle  panels; [Fe/H] - right  panels) 
  derived in this work with the ones that are available in \textit{exoarch} (top row  - red circles), \textit{exo.eu} (middle row  -  green 
  triangles), and \textit{exo.org} (bottom row  - orange squares). The comparison shows general good consistency, but reveals some outliers that are discussed in the text.}
  \label{Fig_comp}
\end{figure*}

\subsection{Comparison with literature}

To perform a comparison with values of spectroscopic parameters derived in other works we used the following 
online catalogues to collect the data: the Extrasolar Planets Encyclopaedia\footnote{\url{http://exoplanet.eu}} 
\citep[hereafter \textit{exo.eu};][]{Schneider-2011}, the exoplanets.org\footnote{\url{http://exoplanets.org}} \citep[here after \textit{exo.org};][]{Wright-2011}, and 
the NASA Exoplanet Archive\footnote{\url{http://exoplanetarchive.ipac.caltech.edu/}}(here after \textit{exoarch}).

Figure \ref{Fig_comp} shows the comparisons of the effective temperature, surface gravity, and 
metallicity  for the stars in common between each of the catalogues and our sample\footnote{For very few cases where the error 
on log g was not available in \textit{exo.org} we considered the maximum value of error in the catalogue for the stars in common (0.7 dex)}. For 
both the \textit{exo.eu} and \textit{exo.org} 
we were able to find parameters for almost all the stars in our sample, while for the \textit{exoarch} we have fewer stars with parameters 
available for the comparison. We note that for the case of \textit{exo.eu} we do not have the stellar log g available.

The comparisons are generally quite consistent for the temperatures, surface gravities, and metallicities, with 
standard deviations of the differences around 100 K, 0.25 dex, and 0.1 dex, respectively. The significant high dispersion certainly comes from the 
heterogeneous nature of the literature compilations. 

There are three stars in the comparison with significant large differences in effective temperature (see Fig. \ref{Fig_comp}).
These are the cases of \object{42 Dra}, \object{HAT-P-14}, and \object{HD 118203}. Interestingly, these stars have very different temperatures 
covering different spectral types. 

\paragraph{\object{42 Dra}:} This is a K giant star and in both the \textit{exo.eu} and \textit{exoarch} catalogues the provided 
temperature is 4200 K \citep[][]{Dollinger-2009}, which is a difference of 
about 250 K with the one derived in this work (4452 K). Using  the VizieR database\footnote{\url{www.vizier.u-strasbg.fr/viz-bin/VizieR}}, 
we found a dispersion of temperatures for this star between 4200 K to 4500 K. We note that 
most of the temperatures in VizieR for this star fall in the range 4400-4500 K, which is in good agreement with our measurement.

\paragraph{\object{HAT-P-14}:} This is a F dwarf star, and actually the hottest star in the sample, appearing in the top-right corner of the left 
panels in Fig. \ref{Fig_comp}. The difference is again nearly 250 K (6845 K being our value against the 6600 K found in the online databases). 
Making a new census using VizieR to find additional temperatures for this star, 
we found only a few values ranging from 6427 K to 6671 K. Our value is still compatible with the higher values found when considering the errors.

\paragraph{\object{HD 118203}:} This is a star much more similar to our Sun. However, when comparing it with the exoplanet databases the difference is surprisingly large, 
almost 300 K (5890 K being our value against the 5600 K found in the online databases). The values that we found in the literature, again using  VizieR, 
vary from 5600 K up to 5910 K. Given  that our analysis is actually a differential analysis relative to the 
Sun \citep[][]{Sousa-2008, Sousa-2014}, we are very confident in our temperature derivation for this star.

For the surface gravity the comparisons are quite consistent, with larger dispersion for the giant stars (surface gravity lower than 2.5 dex), but without 
a clear outlier that deserves to be discussed. The good consistency in surface gravity for giants is in line with our latest results \citep[][]{Mortier-2014}.

The dispersion in metallicity is quite significant, but \object{HD 139357} clearly stands out as 
an outlier showing a difference of more than 0.3 dex between the values that we found on \textit{exo.eu} and \textit{exo.org} (-0.13 dex in both cases) and 
the value derived in this work (0.19 dex). As before, using VizieR to find additional values in the literature, we found values ranging from 
 -0.13 dex up to as much as 0.34 dex. Such differences in metallicity might indeed  significantly affect 
the stellar mass estimation and consequently the mass of the planet (see Sect. 4.4 for a deeper discussion of this particular case).

\begin{figure*}[ht]
  \centering
  \includegraphics[width=16cm]{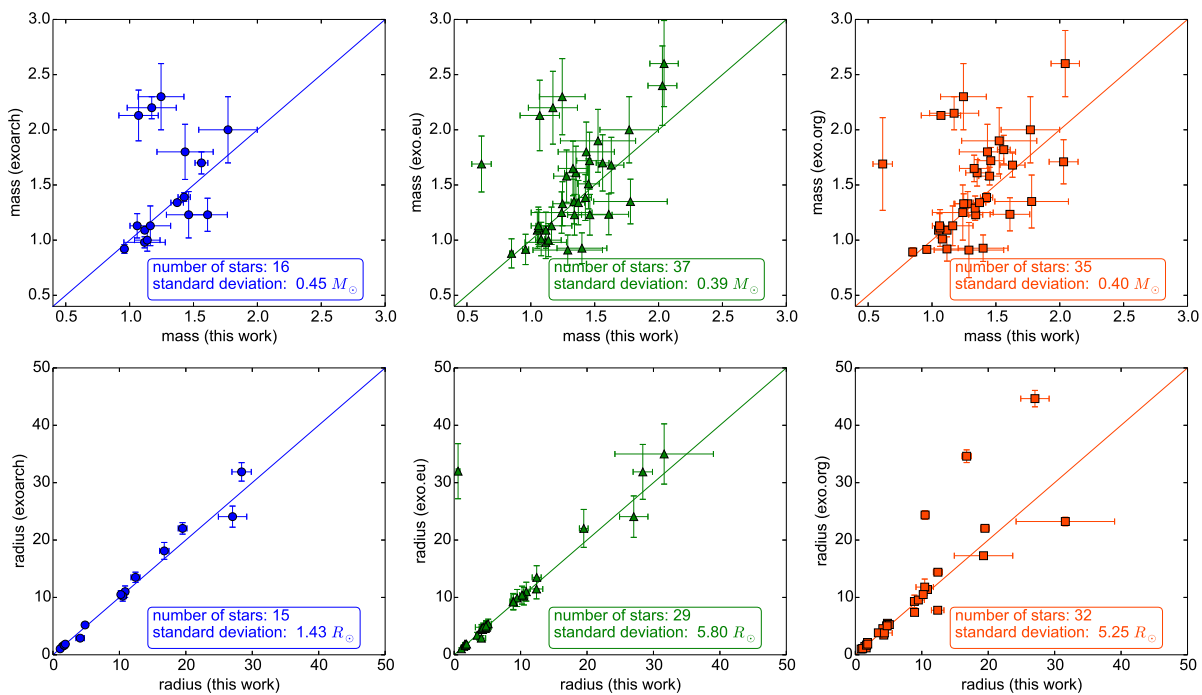}
  \caption{Comparison of the stellar masses (top row) and stellar radius (bottom row) 
  estimated in this work with the ones that are available in \textit{exoarch} (left column  - red circles), \textit{exo.eu} (middle column  -  green 
  triangles), and \textit{exo.org} (right column  - orange squares). The comparison reveals some extreme outliers that are discussed in detail in the text.}
  \label{Fig_comp_mass}
\end{figure*}


\section{Stellar masses}

\subsection{Methods used to estimate the stellar mass}

To estimate the mass for these planet host stars in a homogeneous way we  follow the same procedure that was used 
in SWEET-CAT \citep[see][]{Santos-2013}. Because most of 
the planet hosts analyzed here are evolved stars, we used the web interface for the Bayesian estimation of stellar parameters based on the PARSEC 
isochrones \citep[][]{daSilva-2006, Bressan-2012}\footnote{\url{http://stev.oapd.inaf.it/cgi-bin/param}}. This web interface also derives 
an estimation for the stellar radius and the age. To use this tool we need the parallax values for the stars. These were compiled from the 
SIMBAD astronomical database whenever they were present. For the cases where the parallax value is not available, we computed the spectroscopic 
parallax using our derived spectroscopic stellar parameters (for more details on the iterative method see \citet[][]{Sousa-2011}).

\begin{figure}
  \centering
  \includegraphics[width=8cm]{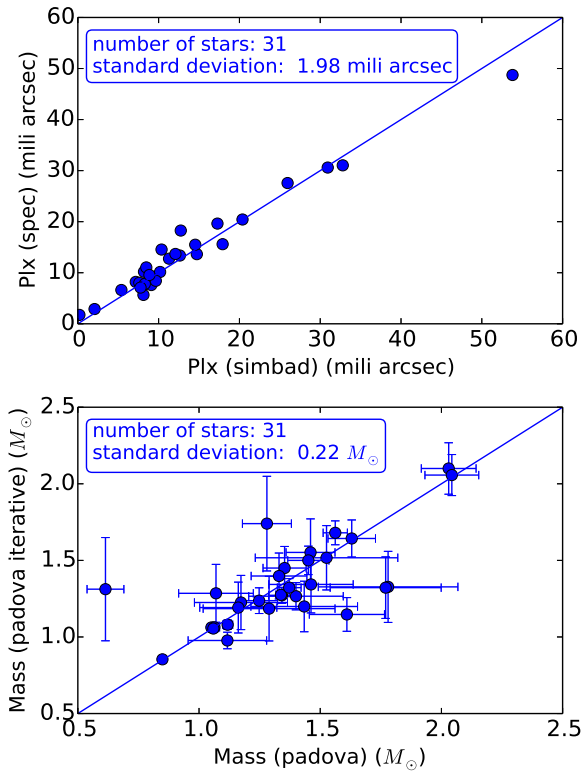}
  \caption{Top panel: Comparison between the parallax values found on SIMBAD and the derived spectroscopic parallax values. 
           Bottom panel: Comparison between the estimated stellar mass using the different parallax values. These panels show that 
           the iterative method to derive paralax/mass without the distance information works fairly well.}
  \label{Fig_comp_plx}
\end{figure}

To double check the results from this iterative method we computed the spectroscopic parallax (and iterative mass) for all the stars 
analyzed in this work. The comparison for the stars that have parallax present in SIMBAD can be seen in Fig. \ref{Fig_comp_plx}. The top 
panel shows a direct comparison between the parallax values showing a dispersion with a standard deviation of 2 milliarcseconds. The bottom 
plot of the same figure shows the comparison of the masses computed using the different parallax values. As already shown 
in \citet[][]{Sousa-2011}, the comparison is in general agreement with a significant large dispersion on the masses. The 
standard deviation here has a value of 0.22 $M_\sun$. We note that a few stars (\object{4 UMa}, \object{HD 116029}, \object{HD 139357}, 
\object{HD 173416}, and \object{HD 240237}) show a significant difference in the mass determinations by the two methods, thus heavily 
contributing to the large dispersion. If we remove these five stars from the 
comparison, then the standard deviation drops to 0.09 solar masses, which is compatible with the previously observed result \citep[][]{Sousa-2011}.

The presence of the outliers observed in Fig. \ref{Fig_comp_plx} already alert us to be cautious with the use of fast stellar mass estimations like 
the tool that we are using here. For most of the cases we are able to derive masses within $\sim$ 10\% accuracy, although there are some cases where the mass estimation has larger errors.
We note that all  five outlier stars in the bottom panel present  larger errors than   the rest of the stars. This issue should be 
considered very important, especially if we keep in mind that the inaccuracy on the stellar mass is  transferred directly to the 
determination of the planetary masses ($M_p \propto M_{*}^{2/3}$, where $M_p$ and $M_{*}$ are the planetary 
and stellar masses, respectively). In Sect. 4.3 we  discuss some examples where we bring up this issue again.

Table \ref{tab3} contains the results that we obtained for the stellar mass, radius, age, as well as the visual magnitude and the 
parallax values used as input for the web interface tool. To keep consistency in the catalogue, for the few  unevolved stars (with log g $\ge$ 4) 
we also derived the mass using the calibration of \citet[][]{Torres-2010}. For this calibration we used the spectroscopic parameters derived previously 
and applied the small correction presented in Eq. (1) of \citet[][]{Santos-2013}. These values are also presented in Table \ref{tab3}, and as expected, they 
are in good agreement with those derived with the web interface.

\subsection{Comparison with literature}

As in the previous section, we used the same online exoplanet databases to get values of masses and radius from the literature and compare 
them with our estimations. Figure \ref{Fig_comp_mass} shows the result of the direct comparison of these fundamental parameters for the stars in common 
between our sample and each online database. For the cases where the errors for the mass or radius were not available in the online databases, we chose 
to adopt a 15\% uncertainty, which is a conservative value when compared with the available values. The different comparisons of the masses are very 
similar for most of the stars, showing again that the online databases  use the 
same sources for the majority of the stars in common. In the top 
panels we can see a significant large dispersion on the mass. There are five planet hosts that present significant offsets, but in this case all in 
the same direction. They are \object{14 And}, \object{HD 17092}, \object{HD 180314}, \object{HD 240237}, and \object{Omi CrB}. We note that these stars are not 
the same outliers  that appeared before when discussing the parallax values, except the case of \object{HD 240237}. For this star we note that using 
the iterative method we found a mass of 1.34 $M_\sun$, which is in between the 
values in Fig. \ref{Fig_comp_mass} (0.61 $M_\sun$ and 1.69 $M_\sun$). In the next section we   discuss in more detail the effect of our estimated mass values for these five 
outlier stars and the effect that it has on the respective planetary masses. Looking back at the top panels, if we neglect the three to five most extreme outlier stars 
present in each plot, we then obtain a better agreement between the masses that have  less than half the original standard deviation ($\sim$ 0.2 $M_\sun$).

Because the stellar radius is also fundamental for the characterization of the transiting planet hosts, we present in the bottom panels of 
Fig. \ref{Fig_comp_mass} the comparison of the stellar radius that we find in the same exoplanet databases. The results 
are generally in good agreement with only very few outliers. In the comparison with the values that we found in 
\textit{exo.eu} the very strong outlier is again the star \object{HD 240237}. Looking  again at the results obtained with the iterative 
method, we obtained a radius of 17.9 $R_\sun$ which is also in 
between the values in the plot (0.59 $R_\sun$ and 31 $R_\sun$). The three outliers present in the comparison with the radius for stars in common with 
\textit{exo.org} are the planet hosts \object{11 UMi}, \object{4 UMa,} and \object{HD 32518}.

\subsection{Impact of planetary masses}

In this section we want to address the impact of the derived stellar mass on the planetary mass (here mass is defined as m$\sin i$ since most cases in 
this work are radial velocity detections, where $i$ is the orbital inclination to the line of sight). We  
exemplify this effect using the extreme outlier cases described in the previous section.

\paragraph{\object{14 And}:}
\citet[][]{Sato-2008} have  detected a giant planet with 4.8 $M_{jup}$ around \object{14 And}, assuming a stellar mass of 2.2 $M_\sun$. Using our mass determination 
for this star, we can recalculate the planetary mass assuming the relation already mentioned  ($M_p \propto M_{*}^{2/3}$). Using this 
relation and our mass determination (1.17 $M_\sun$) we reach  a planetary mass of only 2.22 $M_{jup}$, which is a change of more than 100\%. Recently, also 
for this planet host, \citet[][]{Ligi-2012} have reviewed the mass of this planet using interferometry. With a radius derived from interferometry 
of 12.82 $\pm$ 0.32 $R_\sun$, these authors then used the surface gravity derived 
in \citet[][]{Sato-2008} (2.63 dex) and the gravitational acceleration relation to derive a stellar mass reaching a value of 2.6 $\pm$ 0.42 $M_\sun$. If we do the 
same exercise using the stellar radius from interferometry together with our determination of surface gravity (2.44 dex) we reach a stellar mass 
of only 1.68 $M_\sun$, which then translates into a planetary mass of 3.18 $M_{jup}$, which still represents a difference of $\sim$ 66\%.

\paragraph{\object{HD 17092}:} \citet[][]{Niedzielski-2007} reported the presence of a 4.6 $M_{jup}$ planet around \object{HD 17092}. A mass of 2.3 $M_\sun$ for 
the stellar host was adopted by these authors to derive the mass of the companion. Using our mass determination (1.25 $M_\sun$) we reach to a planetary 
mass of 2.2 $M_{jup}$. Again more than a 100\% difference. 

\paragraph{\object{HD 180314}:} This star represents another interesting case. \citet[][]{Sato-2010} detected a substellar companion around \object{HD 180314} 
with 22 $M_{jup}$. The estimated stellar mass in this work was 2.6 $M_\sun$. If we use our stellar mass estimation of 2.04 $M_\sun$ we reach a less massive 
companion of 15.8 $M_{jup}$.

\paragraph{\object{HD 240237}:} This object was  mentioned before because we found very different values for both its mass and radius. 
Doing the same exercise using the values reported in \citet[][]{Gettel-2012} (a 5.3 $M_{jup}$ giant planet derived with a stellar mass of 1.69 $M_\sun$) and using our 
mass determination of only 0.61 $M_\sun$ we obtain a planetary mass of only 1.53 $M_{jup}$.

\paragraph{\object{omi CrB}:} \citet[][]{Sato-2012} reported a 1.5 $M_{jup}$ giant planet orbiting around \object{omi CrB}, assuming a stellar mass of 2.13 $M_\sun$. With 
our mass determination of 1.07 $M_\sun$ we put the planet at a sub-Jupiter mass class with only 0.65 $M_{jup}$.

Our goal with these examples is not to state that our mass estimations are the best ones, defining therefore new, precise, and better planetary masses. Instead, 
we want to show that the estimation of the stellar mass for the planet hosts should be very well studied, especially in the more 
problematic case of evolved stars. On the one hand, it is known that the determination of spectroscopic stellar parameters can change significantly, in particular 
when using different line lists in the spectroscopic analysis \citep[e.g.,][]{Taylor-2005, Santos-2012, Mortier-2013, Alves-2015}. 
On the other hand, the estimation of the fundamental parameters using stellar modeling can also lead to significant differences and 
interesting discussions such as the ones of \citet[][]{Lloyd-2011, Lloyd-2013} and \citet[][]{Johnson-2013}.
These simple examples presented here show that the different estimations of the stellar mass can  change 
significantly the understanding of the discovered planets. Taking this to extreme examples, it can eventually reclassify the status of stellar companions.

\subsection{Mass and metallicity in \object{HD 139357}}

In the previous examples we showed that by using our mass estimation the planetary masses  all changed to smaller values. We did not have any star that 
would produce the opposite scenario where the mass of the planet would be larger. Here we  present such a case, but for this example the mass estimation 
is closely related with the metallicity of the star. 

In Sect. 3.2. we report that the planet host \object{HD 139357} has a large difference when comparing the metallicity derived in this work with the one that 
we found in the online databases. For the case of \textit{exo.eu}, the values reported for metallicity and temperature are -0.13 dex and 4700 K, respectively, 
derived by \citet[][]{Dollinger-2009}. If we use these values in the web-interface to estimate the stellar mass we obtain a value of 
1.36 $\pm$ 0.18 $M_\sun$, which is in perfect agreement with the value used in the same work (1.35 $\pm$ 0.24 $M_\sun$) and which was used to derive a planetary 
mass of 9.76 $M_{jup}$.

If we go back to our parameters derived for this star, we have a temperature of 4595 K $\pm$ 76, which is compatible with the one derived 
by \citet[][]{Dollinger-2009}. However, as reported in section 3.2. we obtained 0.19 dex for metallicity, corresponding to a large difference
of 0.34 dex. The mass that we report in table \ref{tab3} derived using our spectroscopic values is 1.78 $M_\sun$, which is significantly larger. 
Using this to recompute the planetary mass we obtain a value of 11.7 $M_{jup}$.

To test if this change could be due to the small difference in the derived effective temperature ($\sim$ 100 K) we used our spectroscopic 
parameters, changing only the metallicity, to derive  the stellar mass again. The result is 1.33 $M_\sun$, 
which is very close to the value reported in \citet[][]{Dollinger-2009}. This means that the stellar metallicity determination alone can be responsible 
for a change of 20\% in the planetary mass for this extreme case. This example shows that we can also strongly underestimate a planetary mass in 
opposition with the examples that were described in the previous section.

\section{Summary}

We have collected high resolution and high signal-to-noise spectra with the NARVAL 
spectrograph. We used this data to derive new, precise, and homogeneous spectroscopic parameters 
using our standard analysis (ARES+MOOG) for 37 FGK planet host stars bright in different stages of evolution. 

We also estimated the mass, radius, and age for all the stars in the sample following the same 
procedure as in previous works. The homogeneous parameters derived in this 
work were then included in the SWEET-Cat catalogue making it even more complete for the RV detected 
planets.

We compare our values with the ones available in online databases for exoplanets. The results 
are generally consistent, but we were able to identify some targets for which significant offsets are present, especially 
for  the mass determination of a few giant stars in the sample.

We also discuss the effect of the stellar mass determination on the planetary mass. We show the most extreme cases for 
which the planetary masses may be overestimated. In addition, we also spotted an example with the 
opposite trend. In this case the source for the underestimation of the planetary mass is the stellar 
metallicity determination. We show with clear examples that it is fundamental to have precise stellar masses, which are not 
easy to obtain. To achieve this, spectroscopic stellar parameters are fundamental in order to constrain the stellar models.

\begin{acknowledgements}

This work is supported by the European Research Council/European 
Community under the FP7 through Starting Grant agreement
number 239953. N.C.S. was supported by FCT through the Investigador FCT 
contract reference IF/00169/2012 and POPH/FSE (EC) by FEDER funding 
through the program Programa Operacional de Factores de Competitividade - COMPETE.
S.G.S, E.D.M, and V.Zh.A. acknowledge the
support from the Funda\c{c}\~ao para a Ci\^encia e Tecnologia, FCT (Portugal) and POPH/FSE (EC), in the
form of the fellowships SFRH/BPD/47611/2008, SFRH/BPD/76606/2011, and
SFRH/BPD/70574/2010. A.M. acknowledges support from the European Union Seventh 
Framework Programme (FP7/2007-2013) under grant agreement number 313014 (ETAEARTH).
G.I. acknowledges financial support from the Spanish
Ministry project MINECO AYA2011-29060. V. N. acknowledges a CNPq/BJT Post-Doctorate fellowship 301186/2014-6 and partial financial support of the INCT INEspa\c{c}o.
This research has made use of the SIMBAD database operated at CDS, Strasbourg, France. The authors acknowledge 
support from OPTICON (EU-FP7 Grant number 312430)
\end{acknowledgements}

\bibliographystyle{bibtex/aa}
\bibliography{sousa_bibliography}

\begin{appendix}

\section{Spectroscopic parameters}

\begin{table*}[ht]
\caption[]{Spectroscopic parameters derived with ARES+MOOG.}
\begin{center}

\begin{tabular}{rccccccc}
\hline
\hline
Star ID     & T$_{\mathrm{eff}}$    & $\log{g}_{spec}$    & $\xi_{\mathrm{t}}$ & \multicolumn{1}{c}{[Fe/H]} & N(\ion{Fe}{i}, \ion{Fe}{ii}) & $\sigma$(\ion{Fe}{i}, \ion{Fe}{ii})   & Transiting planet\\
            & [K]                   & [cm\,s$^{-2}$]      &  [km\,s$^{-1}$]    &     [dex]                  &                              & [dex]                                & [Yes/No] \\
\hline

\object{11 UMi}    & 4255\ $\pm$\  88 & 1.80\ $\pm$\ 0.26 & 1.79\ $\pm$\ 0.08  &  -0.13\ $\pm$\ 0.04        & 113, 15                      & 0.18, 0.30 & No \\
\object{14 And}    & 4709\ $\pm$\  37 & 2.44\ $\pm$\ 0.12 & 1.51\ $\pm$\ 0.03  &  -0.29\ $\pm$\ 0.03        & 120, 15                      & 0.09, 0.20 & No \\
\object{42 Dra}    & 4452\ $\pm$\  42 & 2.09\ $\pm$\ 0.16 & 1.50\ $\pm$\ 0.04  &  -0.41\ $\pm$\ 0.03        & 117, 15                      & 0.10, 0.29 & No \\
\object{4 Uma}     & 4531\ $\pm$\  52 & 2.32\ $\pm$\ 0.14 & 1.54\ $\pm$\ 0.04  &  -0.19\ $\pm$\ 0.03        & 118, 15                      & 0.11, 0.20 & No \\
\object{6 Lyn}     & 5022\ $\pm$\  28 & 3.15\ $\pm$\ 0.13 & 1.16\ $\pm$\ 0.03  &  -0.10\ $\pm$\ 0.02        & 118, 14                      & 0.06, 0.23 & No \\
\object{gam01 Leo} & 4395\ $\pm$\  48 & 1.66\ $\pm$\ 0.13 & 1.67\ $\pm$\ 0.04  &  -0.47\ $\pm$\ 0.03        & 118, 15                      & 0.12, 0.21 & No \\
\object{HAT-P-14}  & 6845\ $\pm$\ 108 & 4.62\ $\pm$\ 0.05 & 1.92\ $\pm$\ 0.15  &   0.17\ $\pm$\ 0.07        & 215, 34                      & 0.19, 0.11 & Yes\\
\object{HAT-P-22}  & 5351\ $\pm$\  57 & 4.22\ $\pm$\ 0.11 & 0.89\ $\pm$\ 0.10  &   0.28\ $\pm$\ 0.05        & 256, 33                      & 0.13, 0.22 & Yes\\
\object{HD 100655} & 4891\ $\pm$\  46 & 2.79\ $\pm$\ 0.11 & 1.37\ $\pm$\ 0.04  &   0.07\ $\pm$\ 0.03        & 118, 15                      & 0.10, 0.17 & No \\
\object{HD 102956} & 5010\ $\pm$\  37 & 3.31\ $\pm$\ 0.08 & 1.12\ $\pm$\ 0.04  &   0.12\ $\pm$\ 0.03        & 117, 15                      & 0.08, 0.09 & No \\
\object{HD 116029} & 4819\ $\pm$\  54 & 2.98\ $\pm$\ 0.12 & 1.03\ $\pm$\ 0.06  &   0.09\ $\pm$\ 0.04        & 118, 15                      & 0.11, 0.16 & No \\
\object{HD 118203} & 5890\ $\pm$\  41 & 4.10\ $\pm$\ 0.06 & 1.25\ $\pm$\ 0.04  &   0.30\ $\pm$\ 0.03        & 252, 33                      & 0.10, 0.13 & No \\
(1)                & 5910\ $\pm$\  35 & 4.18\ $\pm$\ 0.07 & 1.34\ $\pm$\ 0.04  &   0.25\ $\pm$\ 0.03   & - & -  & -  \\
\object{HD 131496} & 4886\ $\pm$\  45 & 3.16\ $\pm$\ 0.11 & 1.16\ $\pm$\ 0.05  &   0.12\ $\pm$\ 0.03        & 119, 15                      & 0.09, 0.15 & No \\
\object{HD 132406} & 5766\ $\pm$\  23 & 4.19\ $\pm$\ 0.03 & 1.05\ $\pm$\ 0.03  &   0.14\ $\pm$\ 0.02        & 253, 35                      & 0.06, 0.06 & No \\
\object{HD 132563B}& 6073\ $\pm$\  51 & 4.42\ $\pm$\ 0.05 & 1.02\ $\pm$\ 0.07  &  -0.06\ $\pm$\ 0.04        & 246, 35                      & 0.11, 0.11 & No \\
\object{HD 136418} & 4979\ $\pm$\  44 & 3.43\ $\pm$\ 0.08 & 1.03\ $\pm$\ 0.05  &  -0.09\ $\pm$\ 0.03        & 118, 14                      & 0.09, 0.08 & No \\
\object{HD 139357} & 4595\ $\pm$\  76 & 2.54\ $\pm$\ 0.20 & 1.61\ $\pm$\ 0.08  &   0.19\ $\pm$\ 0.05        & 117, 15                      & 0.16, 0.31 & No \\
\object{HD 145457} & 4829\ $\pm$\  41 & 2.67\ $\pm$\ 0.10 & 1.48\ $\pm$\ 0.03  &  -0.13\ $\pm$\ 0.03        & 119, 15                      & 0.09, 0.16 & No \\
\object{HD 152581} & 5095\ $\pm$\  23 & 3.31\ $\pm$\ 0.05 & 1.07\ $\pm$\ 0.02  &  -0.30\ $\pm$\ 0.02        & 119, 13                      & 0.05, 0.07 & No \\
\object{HD 154345} & 5442\ $\pm$\  30 & 4.39\ $\pm$\ 0.04 & 0.76\ $\pm$\ 0.06  &  -0.13\ $\pm$\ 0.02        & 256, 33                      & 0.07, 0.10 & No \\
\object{HD 158038} & 4822\ $\pm$\  64 & 3.06\ $\pm$\ 0.16 & 1.11\ $\pm$\ 0.07  &   0.16\ $\pm$\ 0.05        & 119, 15                      & 0.13, 0.23 & No \\
\object{HD 16175}  & 6022\ $\pm$\  34 & 4.21\ $\pm$\ 0.06 & 1.26\ $\pm$\ 0.03  &   0.37\ $\pm$\ 0.03        & 255, 35                      & 0.08, 0.13 & No \\
(1)                & 6030\ $\pm$\  22 & 4.23\ $\pm$\ 0.04 & 1.39\ $\pm$\ 0.02  &   0.32\ $\pm$\ 0.02   & - & -  & -  \\
\object{HD 163607} & 5586\ $\pm$\  29 & 4.05\ $\pm$\ 0.05 & 1.09\ $\pm$\ 0.03  &   0.22\ $\pm$\ 0.02        & 255, 35                      & 0.07, 0.12 & No \\
\object{HD 17092}  & 4596\ $\pm$\  65 & 2.45\ $\pm$\ 0.17 & 1.55\ $\pm$\ 0.06  &   0.05\ $\pm$\ 0.04        & 118, 15                      & 0.14, 0.25 & No \\
\object{HD 173416} & 4783\ $\pm$\  43 & 2.43\ $\pm$\ 0.09 & 1.49\ $\pm$\ 0.03  &  -0.15\ $\pm$\ 0.03        & 118, 14                      & 0.10, 0.10 & No \\
\object{HD 180314} & 4913\ $\pm$\  60 & 2.90\ $\pm$\ 0.17 & 1.45\ $\pm$\ 0.05  &   0.11\ $\pm$\ 0.04        & 119, 15                      & 0.13, 0.28 & No \\
\object{HD 197037} & 6150\ $\pm$\  34 & 4.37\ $\pm$\ 0.04 & 1.11\ $\pm$\ 0.04  &  -0.16\ $\pm$\ 0.03        & 235, 36                      & 0.08, 0.09 & No \\
\object{HD 219415} & 4787\ $\pm$\  53 & 3.22\ $\pm$\ 0.11 & 1.00\ $\pm$\ 0.06  &   0.03\ $\pm$\ 0.03        & 118, 14                      & 0.10, 0.10 & No \\
\object{HD 222155} & 5750\ $\pm$\  23 & 4.00\ $\pm$\ 0.05 & 1.13\ $\pm$\ 0.02  &  -0.09\ $\pm$\ 0.02        & 253, 34                      & 0.06, 0.11 & No \\
(2)                & 5765\ $\pm$\  22 & 4.10\ $\pm$\ 0.13 & 1.22\ $\pm$\ 0.02  &  -0.11\ $\pm$\ 0.05   & - & -  & -  \\
\object{HD 240210} & 4316\ $\pm$\  78 & 1.91\ $\pm$\ 0.21 & 1.76\ $\pm$\ 0.07  &  -0.14\ $\pm$\ 0.03        & 115, 15                      & 0.16, 0.26 & No \\
\object{HD 240237} & 4422\ $\pm$\ 101 & 1.69\ $\pm$\ 0.24 & 2.31\ $\pm$\ 0.08  &  -0.24\ $\pm$\ 0.06        & 111, 14                      & 0.19, 0.30 & No \\
\object{HD 32518}  & 4661\ $\pm$\  53 & 2.41\ $\pm$\ 0.12 & 1.53\ $\pm$\ 0.05  &  -0.10\ $\pm$\ 0.04        & 119, 13                      & 0.12, 0.16 & No \\
\object{HD 96127}  & 4179\ $\pm$\ 110 & 1.59\ $\pm$\ 0.34 & 2.01\ $\pm$\ 0.11  &  -0.29\ $\pm$\ 0.05        & 114, 14                      & 0.23, 0.40 & No \\
\object{HD 99706}  & 4891\ $\pm$\  35 & 3.07\ $\pm$\ 0.08 & 1.15\ $\pm$\ 0.03  &   0.02\ $\pm$\ 0.03        & 119, 15                      & 0.07, 0.12 & No \\
\object{kappa CrB} & 4889\ $\pm$\  48 & 3.20\ $\pm$\ 0.11 & 1.13\ $\pm$\ 0.04  &   0.11\ $\pm$\ 0.03        & 120, 15                      & 0.09, 0.14 & No \\
\object{Kepler-21} & 6269\ $\pm$\  47 & 4.07\ $\pm$\ 0.06 & 1.35\ $\pm$\ 0.05  &   0.03\ $\pm$\ 0.03        & 236, 34                      & 0.10, 0.12 & Yes\\
\object{omi CrB}   & 4792\ $\pm$\  35 & 2.65\ $\pm$\ 0.14 & 1.48\ $\pm$\ 0.03  &  -0.23\ $\pm$\ 0.03        & 119, 15                      & 0.07, 0.25 & No \\

\hline
\end{tabular}
 
\end{center}
\tablefoot{$\log{g}_{spec}$ is the spectroscopic surface gravity; $\xi_{\mathrm{t}}$ is the microturbulance speed; N(\ion{Fe}{i},\ion{Fe}{ii})
is the number of lines used, and $\sigma$(\ion{Fe}{i}, \ion{Fe}{ii}) is the dispersion of the ion abundances in the spectroscopic analysis; \\
(1) - Spectroscopic stellar parameters from SWEET-Cat derived in \citet[][]{Santos-2013}\\
(2) - Spectroscopic stellar parameters from SWEET-Cat derived in \citet[][]{Boisse-2012}\\
}
\label{tab2}
\end{table*}

\section{Stellar masses and radii}

\begin{table*}[t]
\caption[]{Estimations of the stellar masses, radii, and ages for the planet hosts.}
\begin{center}

\begin{tabular}{lccccccc}
\hline
\hline
Star ID         & V mag         & Plx                                   & Mass (Padova)                   & Mass (Torres) & Radius (Padova)               & Age (Padova) \\
                &               & mili arcsec                           & $M_\sun$                        &    $M_\sun$       &    $R_\sun$                       &    Gyr     \\
\hline
11UMi           &       5.02    &        8.19   \ $\pm$\        0.19    &       1.434\ $\pm$\ 0.220    &       -       &       27.032 \ $\pm$\ 2.135   &       3.186\ $\pm$\ 1.521    \\
14And           &       5.22    &       12.63   \ $\pm$\        0.27    &       1.173\ $\pm$\ 0.192    &       -       &       10.859 \ $\pm$\ 0.409   &       5.477\ $\pm$\ 3.089    \\
42Dra           &       4.83    &       10.36   \ $\pm$\        0.20    &       1.117\ $\pm$\ 0.162    &       -       &       19.507 \ $\pm$\ 0.655   &       5.911\ $\pm$\ 3.002    \\
4Uma            &       4.60    &       12.74   \ $\pm$\        0.26    &       1.610\ $\pm$\ 0.155    &       -       &       16.751 \ $\pm$\ 0.704   &       2.080\ $\pm$\ 0.625    \\
6Lyn            &       5.88    &       17.92   \ $\pm$\        0.47    &       1.562\ $\pm$\ 0.050    &       -       &        4.830 \ $\pm$\ 0.164   &       2.353\ $\pm$\ 0.211    \\
gam01Leo        &       2.12    &       25.96   \ $\pm$\        0.83    &       1.462\ $\pm$\ 0.174    &       -       &       28.395 \ $\pm$\ 1.458   &       2.412\ $\pm$\ 0.879    \\
HAT-P-14        &       9.99    &        6.42*                          &       1.427\ $\pm$\ 0.048    &       1.364   &        1.426 \ $\pm$\ 0.078   &       0.431\ $\pm$\ 0.295    \\
HAT-P-22        &       9.76    &        9.87*                          &       0.959\ $\pm$\ 0.024    &       0.985   &        1.084 \ $\pm$\ 0.120   &       9.407\ $\pm$\ 3.130    \\
HD100655        &       6.44    &        8.18   \ $\pm$\        0.50    &       2.030\ $\pm$\ 0.113    &       -       &        8.891 \ $\pm$\ 0.483   &       1.309\ $\pm$\ 0.236    \\
HD102956        &       7.85    &        7.92   \ $\pm$\        0.83    &       1.630\ $\pm$\ 0.098    &       -       &        4.275 \ $\pm$\ 0.453   &       2.328\ $\pm$\ 0.418    \\
HD116029        &       7.89    &        8.12   \ $\pm$\        0.65    &       1.280\ $\pm$\ 0.101    &       -       &        4.693 \ $\pm$\ 0.406   &       4.855\ $\pm$\ 1.307    \\
HD118203        &       8.06    &       11.29   \ $\pm$\        0.70    &       1.342\ $\pm$\ 0.055    &       1.279   &        1.829 \ $\pm$\ 0.119   &       3.578\ $\pm$\ 0.492    \\
HD131496        &       7.80    &        9.09   \ $\pm$\        0.78    &       1.353\ $\pm$\ 0.089    &       -       &        4.132 \ $\pm$\ 0.368   &       4.060\ $\pm$\ 0.821    \\
HD132406        &       8.45    &       14.73   \ $\pm$\        0.61    &       1.051\ $\pm$\ 0.014    &       1.105   &        1.245 \ $\pm$\ 0.055   &       7.187\ $\pm$\ 0.660    \\
HD132563B       &       9.61    &        9.12*                          &       1.081\ $\pm$\ 0.029    &       1.049   &        1.092 \ $\pm$\ 0.080   &       2.310\ $\pm$\ 1.886    \\
HD136418        &       7.88    &       10.18   \ $\pm$\        0.58    &       1.248\ $\pm$\ 0.076    &       1.303   &        3.463 \ $\pm$\ 0.220   &       4.649\ $\pm$\ 0.985    \\
HD139357        &       5.98    &        8.47   \ $\pm$\        0.30    &       1.780\ $\pm$\ 0.288    &       -       &       12.394 \ $\pm$\ 0.941   &       2.028\ $\pm$\ 0.937    \\
HD145457        &       6.57    &        7.98   \ $\pm$\        0.45    &       1.526\ $\pm$\ 0.294    &       -       &        9.483 \ $\pm$\ 0.907   &       2.605\ $\pm$\ 1.608    \\
HD152581        &       8.35    &        5.39   \ $\pm$\        0.96    &       1.400\ $\pm$\ 0.195    &       -       &        4.370 \ $\pm$\ 1.177   &       2.850\ $\pm$\ 1.112    \\
HD154345        &       6.74    &       53.80   \ $\pm$\        0.32    &       0.849\ $\pm$\ 0.015    &       0.862   &        0.869 \ $\pm$\ 0.008   &       9.527\ $\pm$\ 1.662    \\
HD158038        &       7.46    &        9.65   \ $\pm$\        0.74    &       1.330\ $\pm$\ 0.115    &       -       &        4.858 \ $\pm$\ 0.418   &       4.433\ $\pm$\ 1.223    \\
HD16175         &       7.28    &       17.28   \ $\pm$\        0.67    &       1.338\ $\pm$\ 0.022    &       1.292   &        1.633 \ $\pm$\ 0.070   &       2.802\ $\pm$\ 0.162    \\
HD163607        &       7.98    &       14.53   \ $\pm$\        0.46    &       1.118\ $\pm$\ 0.021    &       1.146   &        1.695 \ $\pm$\ 0.063   &       7.581\ $\pm$\ 0.475    \\
HD17092         &       7.73    &        4.59*                          &       1.246\ $\pm$\ 0.179    &       -       &       10.439 \ $\pm$\ 1.310   &       5.580\ $\pm$\ 2.669    \\
HD173416        &       6.06    &        7.17   \ $\pm$\        0.28    &       1.770\ $\pm$\ 0.229    &       -       &       12.422 \ $\pm$\ 0.674   &       1.730\ $\pm$\ 0.653    \\
HD180314        &       6.61    &        7.61   \ $\pm$\        0.39    &       2.043\ $\pm$\ 0.111    &       -       &        8.911 \ $\pm$\ 0.434   &       1.295\ $\pm$\ 0.234    \\
HD197037        &       6.81    &       30.93   \ $\pm$\        0.38    &       1.063\ $\pm$\ 0.022    &       1.059   &        1.105 \ $\pm$\ 0.023   &       3.408\ $\pm$\ 0.924    \\
HD219415        &       8.91    &        5.89*                          &       1.138\ $\pm$\ 0.100    &       -       &        4.098 \ $\pm$\ 0.640   &       7.233\ $\pm$\ 2.295    \\
HD222155        &       7.12    &       20.38   \ $\pm$\        0.62    &       1.059\ $\pm$\ 0.018    &       1.135   &        1.685 \ $\pm$\ 0.059   &       7.854\ $\pm$\ 0.407    \\
HD240210        &       8.27    &        2.41*                          &       1.241\ $\pm$\ 0.238    &       -       &       19.293 \ $\pm$\ 4.399   &       5.085\ $\pm$\ 3.089    \\
HD240237        &       8.17    &        0.19   \ $\pm$\        0.72    &       0.614\ $\pm$\ 0.076    &       -       &        0.587 \ $\pm$\ 0.274   &       4.420\ $\pm$\ 4.007    \\
HD32518         &       6.42    &        8.29   \ $\pm$\        0.58    &       1.162\ $\pm$\ 0.159    &       -       &       10.499 \ $\pm$\ 0.567   &       6.468\ $\pm$\ 3.058    \\
HD96127         &       7.41    &        2.07   \ $\pm$\        0.58    &       1.289\ $\pm$\ 0.272    &       -       &       31.610 \ $\pm$\ 7.414   &       4.067\ $\pm$\ 2.624    \\
HD99706         &       7.64    &        7.76   \ $\pm$\        0.68    &       1.460\ $\pm$\ 0.101    &       -       &        5.120 \ $\pm$\ 0.465   &       3.085\ $\pm$\ 0.625    \\
kappaCrB        &       4.80    &       32.79   \ $\pm$\        0.21    &       1.451\ $\pm$\ 0.085    &       -       &        4.791 \ $\pm$\ 0.165   &       3.286\ $\pm$\ 0.554    \\
Kepler-21       &       8.25    &        8.86   \ $\pm$\        0.58    &       1.372\ $\pm$\ 0.049    &       1.359   &        1.863 \ $\pm$\ 0.126   &       2.818\ $\pm$\ 0.341    \\
omiCrB          &       5.52    &       12.08   \ $\pm$\        0.44    &       1.070\ $\pm$\ 0.154    &       -       &       10.216 \ $\pm$\ 0.351   &       7.827\ $\pm$\ 3.844    \\
\hline
\end{tabular}
 
\end{center}
\tablefoot{Vmag is the Visual magnitude, Plx is the parallax, both values taken from SIMBAD, except for the values marked with `*' (computed spectroscopic 
parallax); (Padova) refers to the values of Mass, Radius, and the Age obtained through the web-interface tool; (Torres) is the mass 
estimation using on the calibration of \citet[][]{Torres-2010}.
}
\label{tab3}
\end{table*}

\end{appendix}

\end{document}